\documentstyle[12pt,epsf,epsfig]{article}
\def\beq{\begin{equation}}
\def\eeq{\end{equation}}
\def\bea{\begin{eqnarray}}
\def\eea{\end{eqnarray}}
\def\bq{\begin{quote}}
\def\eq{\end{quote}}

\def\gappeq{\mathrel{\rlap {\raise.5ex\hbox{$>$}}
{\lower.5ex\hbox{$\sim$}}}}

\def\lappeq{\mathrel{\rlap{\raise.5ex\hbox{$<$}}
{\lower.5ex\hbox{$\sim$}}}}

\def\bigR{\mbox{\boldmath$R$}}

\begin{document}
\pagestyle{empty}
\begin{flushright}
{CERN-TH/98-203}
\end{flushright}
\vspace*{5mm}
\begin{center}
{\bf SUMMARY OF DIS~98}\\
\vspace*{1cm} 
{\bf John ELLIS} \\
\vspace{0.3cm}
Theoretical Physics Division, CERN \\
CH - 1211 Geneva 23 \\
\vspace*{2cm}  
{\bf ABSTRACT} \\ \end{center}
\vspace*{5mm}
\noindent
A personal selection is made of some of the hot
topics debated at this conference, including examples of using 
our knowledge of QCD to
make electroweak measurements, structure functions at low $x$ in the
light of the corrections to the leading BFKL behaviour recently
calculated, diffraction, the existence of one or more Pomerons and
whether it/they may have a well-defined structure function, some
issues in hadronic final states and polarized structure functions,
and the interesting events at large $Q^2$ and $x$ and with isolated
leptons and missing transverse energy. Finally, some of the prospects
for future deep-inelastic scattering facilities are reviewed, and
the interested community encouraged to organize itself to advocate
their approval.\vspace*{1cm}

\begin{center}
{\it Summary Talk Presented at the}\\
{\it Sixth International Workshop on Deep-Inelastic Scattering and QCD}\\
{\it Brussels, April 1998} 
\end{center}
\vspace*{1.5cm}
\vspace*{0.5cm}

\begin{flushleft} CERN-TH/98-203 \\
June 1998
\end{flushleft}
\vfill\eject

\setcounter{page}{1}
\pagestyle{plain}

\section{Why Are We Here?}

	Since there have been many excellent reviews of 
the parallel sessions at this 
conference, my talk is not so much a fair summary, more a set of personal
perspectives. Almost the first question I was asked on arrival in Brussels
was: ``Why are you here?" I turn the question around by asking: ``Why are we
here?" The answers were in fact provided by George Sterman~\cite{Sterman} 
at the end of
DIS~97. We are here to search for new physics -- in his words: ``We must use
the QCD we know well to investigate new physics", and to understand better
QCD and nucleon structure -- in his words: ``Pursue the QCD we do not know
well".

It is worth remembering that the primary motivations for high-energy $ep$
colliders -- first CHEEP in 1978~\cite{CHEEP} and subsequently HERA --
were to search for
new physics. Among the topics proposed for these accelerators were: probing
electroweak neutral currents (now starting at HERA), measuring charged
currents (well on its way), producing the $W^\pm$ and $Z^0$ (apparently
starting) and $ep$ collisions with polarized $e^\pm$ beams (not yet
undertaken at
HERA). The physics interest in the early proposals was focussed on large $Q^2$.

However, as we have all been happy to discover, a funny thing happened on the
way to large $Q^2$ -- in fact several, including low-$x$ structure functions
and diffraction, which generated much of the discussion at this meeting.
Large-$Q^2$ events also generated much discussion at DIS~97, though their
excitement here has been somewhat diminished.

In this talk, I start by reviewing some of the ways in which we may use the QCD
we know well in the searches for new physics. Then I wander through the
garden
of QCD we do not know well, plucking a few flowers that please me, before
turning to large $x$ and $Q^2$ and finally commenting on some future prospects.

\section{Examples of Using QCD}

The understanding of QCD for fixed-target deep-inelastic scattering has now
advanced to a detailed stage, enabling precision electroweak measurements to be
made. We heard at this meeting of a new measurement of
$\sin^2\theta_W$ in deep-inelastic $\nu N$ scattering~\cite{NuTeV}:
\beq
\sin^2\theta_W = 0.2253 \pm 0.0019 \pm 0.0010
\label{one}
\eeq
which can be interpreted as a measure of the $W^\pm$ mass:
\beq
m_W = 80.26 \pm 0.11~{\rm GeV}
\label{two}
\eeq
The fact that the extracted value of $m_W$ depends on both $m_t$ and $m_H$
warns us that this is an indirect measurement, which can be compared with the
prediction
\beq
m_W = 80.333 \pm 0.040~{\rm GeV}
\label{three}
\eeq
made on the basis of precision electroweak data from LEP and
SLD~\cite{LEPEWWG}. When combined
with the earlier CCFRR measurement~\cite{CCFRR}, the error in (\ref{two})
is reduced to
0.105 GeV, among the most precise available measurements, and having
significant impact on the global electroweak fit~\cite{LEPEWWG}. 

On the other hand, measurements of the $W^\pm$
mass using charged-current events at HERA~\cite{Howell} still have some
way to go:
\beq
m_W = 78.6^{+2.5 + 3.3}_{-2.9 - 3.0}~{\rm GeV}
\label{four}
\eeq
Will HERA at high luminosity eventually be able to impact significantly the
global fit to electroweak parameters?
Including the NuTeV result, the latest global fit yields~\cite{LEPEWWG}
\beq
m_H = 66^{+74}_{-39}~{\rm GeV}
\label{five}
\eeq
with a nominal 95 \% confidence-level upper limit of 215 GeV, if no allowance
is made for possible systematic errors. It is striking that the most likely
range for the Higgs mass is that within reach of LEP searches! Unfortunately,
the Higgs production cross section at HERA is not very encouraging

Another example of the use of QCD is in predicting parton-parton luminosity
functions for the Tevatron and the LHC. For example, the most recent analysis
of data from HERA and elsewhere enable the gluon-gluon luminosity to be fixed
with a precision~\cite{fits}
\beq
\Delta \left(\tau {dL^{gg}\over d \tau}\right) < 10 \%
\label{six}
\eeq for $\sqrt{\tau} \sim 10^{-2}$ as relevant for Higgs production at the
LHC, if the estimate (\ref{five}) is correct. Likewise, the uncertainty in the
gluon-quark luminosity~\cite{fits}
\beq
\Delta \left(\tau {dL^{gq}\over d \tau}\right) < 10 \%
\label{seven}
\eeq
for $\sqrt{\tau}\sim 10^{-1}$ as relevant single-top production at the
Tevatron. Thus structure function measurements enable the discovery prospects
of hadron-hadron colliders to be assessed reliably.

\section{Structure Functions at Low $x$}

These certainly appear in the category that we do not know well. The
metaphysical question that awaits a definitive answer is: to
BFKL~\cite{BFKL} or not to
BFKL~? that is, are the structure functions dominated by diffusion in $k_T$,
without strong ordering, in which case one should resum $\Sigma (\alpha_s \ln
{1\over x})^n$, or does the DGLAP ordering in $k_T$ still dominate, in which case
one should resum $\Sigma (\alpha_s \ln Q^2)^n$~? In the former case, one has
the following generic high-energy cross-section formula~\cite{Mueller}:
\beq
\sigma_{\rm AB} \sim \int {e^{E(Q_1,Q_2,Y)}\over \sqrt{4\pi D Y}}~:~~D =
{7\alpha_s\over 2\pi}~N_c~ \xi(3)
\label{eight}
\eeq
with the leading high-energy behaviour of the kernel $E$ for fixed scales
$Q_1,Q_2$ as $Y \rightarrow\infty$ given by the BFKL Pomeron intercept:
\beq
\alpha_P (Q_1Q_2) = 1 + {4N_c \ln 2\over \pi} ~~\alpha_s
\label{nine}
\eeq
After heroic calculations by Fadin and Lipatov~\cite{FL}, supported by
Camici, Ciafaloni~\cite{CC}
and others~\cite{KopPesch,others}, we now know the leading corrections to
the
kernel:
\beq
E(Q_1,Q_2,Y) \simeq [\alpha_P (Q_1Q_2) -1 ]~~~[1-6.5 \alpha_s]~Y +
\left({1\over 5} \alpha^5_s\right) Y^3 + {\ln^2 Q_1/Q_2\over\Delta DY}
\label{ten}
\eeq
The factor $[1-6.5 \alpha_s]$ in the first term indicates that the
leading-order Pomeron intercept (\ref{nine}) is strongly modified \footnote{The
second term in (\ref{ten}), which breaks na\"\i ve Regge behaviour, is 
estimated to be small in
accessible kinematic ranges~\cite{Mueller}.}.
Does this mean that the strong growth (\ref{nine}),(\ref{ten}) is actually
uncalculable with present techniques~? and if quantitative predictions are
impossible, how relevant is the qualitative physics of BFKL, namely the absence
of strong ordering in $k_T$~?

Na\"\i vely, the first correction in (\ref{ten}) makes the leading $\ln (1/x)$
``hard" Pomeron even softer than the old non-perturbative ``soft" Pomeron,
which is discouraging. However, about a half of the correction is due to
phase-space effects that could in principle be resummed~\cite{CC}. The
fact that
non-leading terms would surely be important had been emphasized
previously~\cite{before}.
Nevertheless, to my mind, there is hope for some quantitative understanding, if
more exact results become available. So far, we have NLO (and leading
$1/N_f$~\cite{Gracey})
calculations for all deep-inelastic moments, and NNLO calculations for a few
non-singlet moments. Can we hope for a full NNLO calculation before the LHC~?

The reaction~\cite{Ball}: ``Now we know why we never saw BFKL" may be too
rapid. Perhaps we
can, if we are clever. Promising avenues for seeing the qualitative BFKL
physics may be provided by the forward jets in deep-inelastic scattering, and
by hadron-hadron collisions, as discussed later. A particularly clean
place to search for BFKL effects is in $\gamma^*\gamma^*$ collisions.
Estimates~\cite{deRoeck}
are that during 1998 each LEP experiment could see between a few and a few dozen
events in the interesting kinematic range, depending whether the ``soft" Pomeron
or leading-order BFKL dominates. The results may be a hot topic for DIS~99.

Despite the mixed fortunes of leading-order BFKL, many fancier theoretical
ideas are circulating. There has been progress in calculating the
three-gluon-exchange kernel for the odderon~\cite{threegluon}: why not
also the four- and N-gluon kernels~?
Progress has also been made in calculating $2\rightarrow 4$ transition
vertices. These and other aspects of BFKL dynamics can be tackled using
conformal-field-theory techniques and SL(2,$\bigR$)
symmetry. There is also a
fascinating equivalence to condensed-matter models of spin chains -- albeit
with a non-compact spin: $\vert \underline{s}^2\vert = s(s+1) < 0$ -- opening
the way to attacks using the theory of integrable systems, the Bethe Ansatz,
etc.~\cite{FK}. In the continuum limit of very many gluons, this spin
chain
becomes
formally equivalent to a non-compact non-linear $\sigma$ model that has
affinities with the theory of black holes in string theory~\cite{EM}. This
is all very
elegant, but does it have any connection with practical reality?

One kinematic range where there may be a discrepancy with na\"\i ve DGLAP $\ln
Q^2$ evolution is at low $x$ and $Q^2$~\cite{Caldwell}: could this be due
to saturation of the
parton density? According to the normal leading-order evolution equations,
\beq
{\partial F_2\over\partial \ln Q^2}~ \propto~ x~ g~(x,Q^2)
\label{eleven}
\eeq
However, if there is saturation at small $k^2_T$: $\sigma_{in}\simeq R^2_0$,
one would find~\cite{Mueller}
\beq
{\partial F_2\over\partial \ln Q^2}~ \propto~ R^2_0 ~Q^2
\label{twelve}
\eeq
which is compatible with the data. However, one should be careful, and exhaust
the conservative options before jumping to conclusions. It has been
pointed out~\cite{MRST}
that one can (more or less) fit the data by varying $g(x,$ low $Q^2$), without
changing significantly $g(x,$ high $Q^2$),
as shown in Fig.~1. However, the required
low-$Q^2$
gluon distribution is valence-like, and one still needs a low-$x$, low-$Q^2$
$\bar qq$ sea that is independent of the gluons. These ideas do not shock me,
and they can perhaps be tested by studying vector-meson
production~\cite{Mueller}, which is
sensitive to the gluon distribution in this kinematic range. Alternative
interpretations might include the NNLO perturbative QCD corrections that remain
to be calculated, and the possible contribution of $F_L$ should be better
understood.

\begin{figure}
\hglue2.5cm
\epsfig{figure=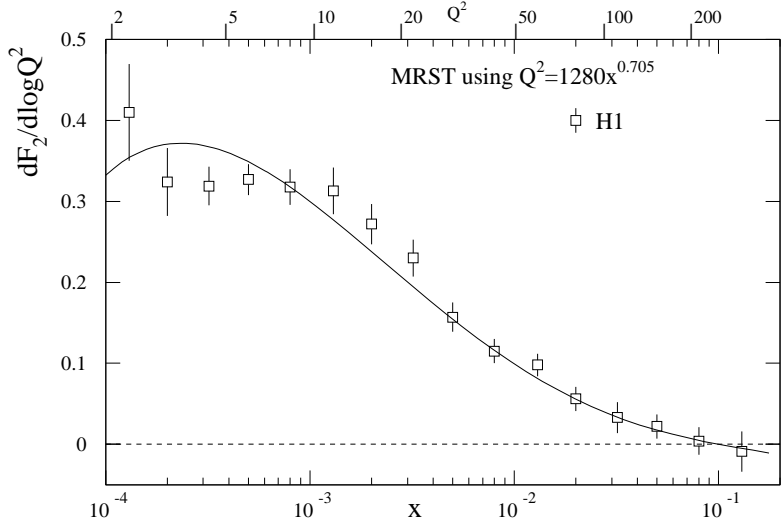,height=7cm}
\caption[]{DGLAP calculation of $dF_2/d \log Q^2$
compared with H1 data~\cite{MRST}.}
\label{fig:une}
\end{figure}

\section{Diffraction}

HERA is now providing us with a wealth of data on diffraction, sharpening the
perennial question: Are there one, two or many Pomerons~?~\cite{DL}  The
``soft" Pomeron
that is assigned responsibility for the rises in the $K^+p, pp, \bar pp, \pi p$
and $\gamma p$ cross sections: $\sigma\sim s^\epsilon$ would have intercept
$\alpha_P(0) = 1 + \epsilon \simeq$ 1.08. Asymptotically, any  rise must
eventually respect the Froissart bound imposed by unitarity: $\sigma < c (\ln
s)^2$. However, the theoretical limiting coefficient $c$ is relatively large, so
this bound may be irrelevant at present energies. There is,
however, evidence that single soft-Pomeron exchange  is not adequate, for
example in single diffraction~\cite{Goul}. According to a simple
factorizing Regge-pole
picture, one would expect
\beq
\sigma_{\rm SD} \sim s^{2\alpha_P(0)-2} \sim s^{2\epsilon}
\label{thirteen}
\eeq
but the data lie far below this estimate, and correspond to an approximately
constant fraction of the total cross section. This observation is suggestive of
saturation effects (multiple soft-Pomeron exchange) already, long before the
Froissart bound enforces it.

This problem of premature saturation can be visualized more clearly in the
impact-parameter picture advocated here by Mueller~\cite{Mueller}.
The $S$ matrix for elastic $pp$ scattering as a function of the impact
parameter $b$ is related to observable cross sections by
\bea
\sigma_{\rm tot} &=& 2 \int d^2b [1-S(b)] \nonumber \\
\sigma_{\rm el} &=& \int t d^2b [1-S(b)]^2 \nonumber \\
\sigma_{\rm inel} &=& \int d^2b [ 1-S^2(b)]
\label{fourteen}
\eea
It has been known since the old ISR days that at high energies $S(b)$ is small as
$b\rightarrow 0$, and that the increasing cross section is due mainly to
expansion in the impact-parameter profile.

The rate at which the cross section for any given process will rise is
determined by the part of the profile which dominates. For example, it has been
suggested that the single-diffraction cross section may be dominated by
intermediate values of $b$ where $0 \ll S(b) \ll 1$. Saturation is reflected in
the importance of multi-Pomeron exchange, in which additional particle
production fills in an erstwhile rapidity gap.

As was also recalled here by Mueller~\cite{Mueller}, it has been
proposed~\cite{impact} that one could probe
the impact parameter profile in aligned-jet production in the reaction
$\gamma^*+p\rightarrow M_x+p$. Defining

\beq
F(x,b) = \int d^2p~e^{ip\cdot b}~\sqrt{{d\sigma_{SD}\over d^2p}}
\label{fifteen}
\eeq
the suggestion is to look at
\beq
\Delta_{\rm eff} \equiv {d~\ln F(x,b)\over d~\ln 1/x}
\label{sixteen}
\eeq
where $x^{-1}$ takes the place of $s$ in high-energy $pp$ scattering. If the
saturation idea is correct, one would expect $\Delta_{\rm eff}$ to be
relatively small when $b\rightarrow 0$.

Large-mass diffraction $\gamma^*+p\rightarrow M_x+p$ can also be used to probe
Pomeron structure in more subtle ways. Let us consider a two-gluon
ladder-exchange model for the Pomeron: then the transverse size of the
$\gamma^*$ state probed is determined by the transverse momentum $k_T$ of the
softest gluon. We may write:
\beq
d\sigma_{SD} \propto {dk^2_T\over Q^2} ~\left[1-S(k_T,b,Y \equiv \ln
(1/x_{P}))\right]^2~d^2 b~dx_{P}
\label{seventeen}
\eeq
In the lowest-order two-gluon model
\beq
1-S \propto x_{P}~~{G(x_{P},k^2_T)\over k^2_T}
\label{eighteen}
\eeq
where $G(x_{P},k^2_T)$ is the gluon density as a function of the
longitudinal momentum fraction $x_{P}$ and the transverse momentum $k_T$.
One might na\"\i vely expect on the basis of (\ref{eighteen}) to find dominance
by low
$k_T\sim\Lambda_{\rm QCD}$. However, as already mentioned, the data on
high-energy hadron scattering suggest that $S\ll 1$ for $b\rightarrow 0$ and
small $k_T$. Hence, and for other reasons~\cite{ER}, the diffractive
$\gamma^*+p\rightarrow M_x+p$ process may get
important contributions from regions where $k_T >\Lambda_{\rm
QCD}$. This
``semihard" Pomeron may lead to faster cross-section growth than the ``soft"
Pomeron.

This discussion indicates that there is just one Pomeron, but that it comes in
many guises. The Pomeron is not a simple Regge pole, as could also be seen from
the previous discussion of BFKL, the historical discussion of cuts and
unitarity, and the rich harvest of data from HERA and elsewhere. The
Pomeron
exists in a multi-parameter space -- these include $b$ and $k_T$ in the
two-gluon approximation alone, but many more parameters would be needed to
characterize multi-gluon exchange. These may be characterized by different
rates of energy growth, and $a~fortiori$ different slopes: the striking new
data~\cite{Jpsi} indicating that the Pomeron is essentially flat in
$\gamma +p\rightarrow
J/\psi + p$ may be particularly helpful in unravelling the two-gluon component. A
complete understanding will require a joint campaign by many experiments: HERA,
the Fermilab Tevatron collider, the LHC, $\cdots$.

What does this discussion tell us about the concept~\cite{IS} of the
``Pomeron structure
function"~? The idea that there may be such universal distributions
$f^{P}_{q,g}(\beta)$ of partons in the Pomeron is very appealing
phenomenologically. However, it is questionable theoretically:
unlike the
$\pi$, there is no nearby, identifiable particle state, and we have
just argued that the Pomeron comes in many guises, so that the concept
of a universal structure function looks implausible.
Moreover, there are experimental
indications from hadron-hadron collisions that factorization into a product of
universal distributions breaks down~\cite{Whitmore}.

On the other hand, the Pomeron structure function idea provides an attractive
description of H1 data~\cite{H194} on the $F_2^{D(3)}$ diffractive
structure function in
electroproduction. These are well described by parton distribution functions
in the Pomeron with a dominant hard gluon at low $Q^2$. This is softened by the
DGLAP equations as $Q^2$ increases, providing a pattern of scaling violations
similar to those seen in the data.

A number of authors have proposed an alternative $\gamma^*$-dissociation
picture~\cite{dissoc}, according to which inelastic diffraction is viewed
as a convolution of
a photon light-cone wave function $\Psi$, an eikonal gluon scattering
kernel, and a proton structure function $\Phi$~\footnote{There are also
related
colour-dipole models~\cite{dipole}.}. This type of model has been used to
predict and to fit
diffractive data from H1 and ZEUS~\cite{dissoc}. There was
considerable
discussion at this meeting of a recent comprehensive fit~\cite{BEKW} with
a phenomenological
model of this type. This model has several components: transverse $\bar qq$
production
$F^T_{\bar qq}$ which is important at medium $\beta$, transverse $\bar qqg$
production $F^T_{\bar qqg}$ which has an importance at small $\beta$ controlled
by an exponent $\gamma$: $F^T_{\bar qqg} \propto (1-\beta)^\gamma$,
longitudinal $\bar qq$ production $\Delta F^L_{\bar qq}$ which is important at
large $\beta$, and a higher-twist contribution $\Delta F^T_{\bar qq}$.

We used this model first to fit the ZEUS 1994 data~\cite{ZEUS94},
as shown in Fig.~2. We
found
a good fit with a
relatively high value of $\gamma$ and with different Pomeron intercepts
$\alpha_P > 1$ for the leading term and the higher-twist contribution $\Delta
F^T_{\bar qq}$. This should not be shocking, given the previous
discussion of the complicated internal structure of the Pomeron. This ZEUS fit
also provided a match to the H1 1994 data that was qualitatively
reasonable~\cite{BEKW},
though not perfect. 
We have
also made fits optimized to the H1 1994 data directly. 
We find two fits, one with
low $\gamma$ (Fig.~3) and one with a higher value (Fig.~4). The solution
with the low value
corresponds to the singular gluon distribution proposed by H1, whereas the
high-$\gamma$ fit resembles more the ZEUS fit. Again, both of the H1 fits are
not too dissimilar from the ZEUS data, but there is a suggestion that ZEUS sees
less $Q^2$ growth.
One may ask whether the two collaborations are 
necessarily measuring exactly the
same thing, in view of their different procedures for event selection.

\begin{figure}
\hglue3.5cm
\epsfig{figure=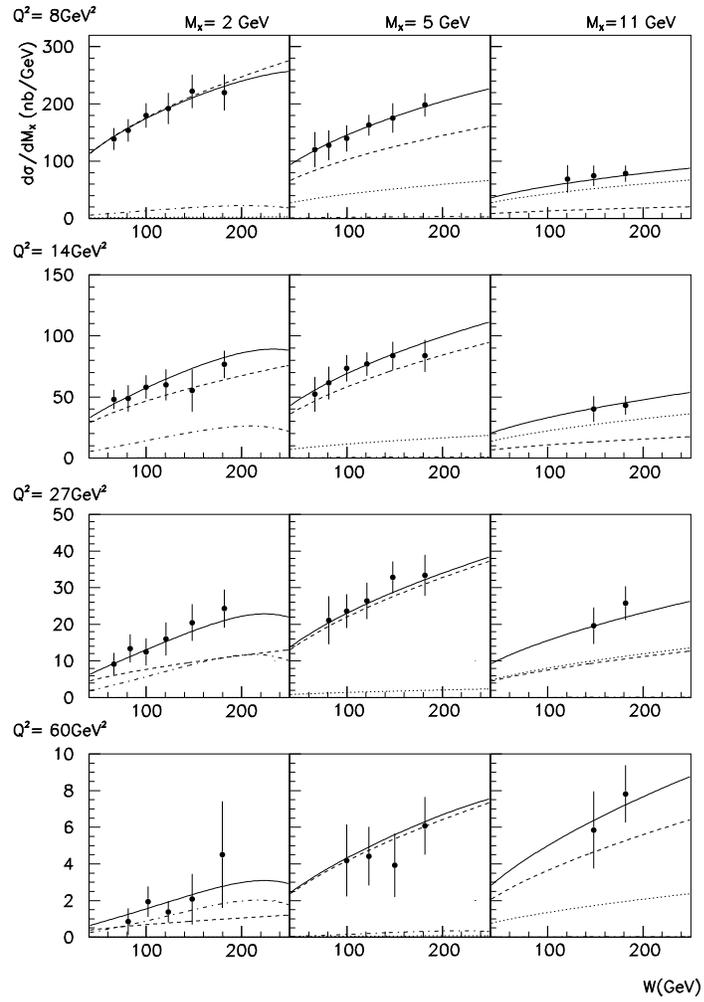,width=10cm}
\caption[]{Fit of two-gluon exchange model to 1994 ZEUS
data on diffraction~\cite{BEKW}.}
\label{fig:deux}
\end{figure}

\begin{figure}
\hglue3.5cm
\epsfig{figure=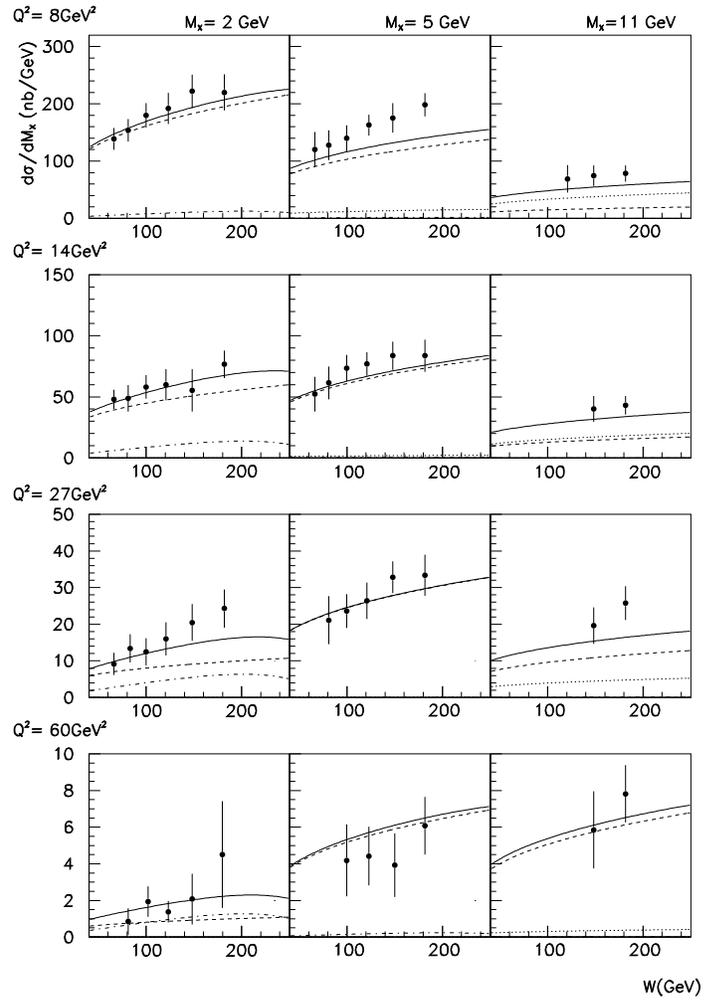,width=10cm}
\caption[]{Fit of two-gluon exchange model to H1 data, 
corresponding to a hard initial gluon distribution in the
Pomeron~\cite{BEKW}.}
\label{fig:trois}
\end{figure}

\begin{figure}
\hglue3.5cm
\epsfig{figure=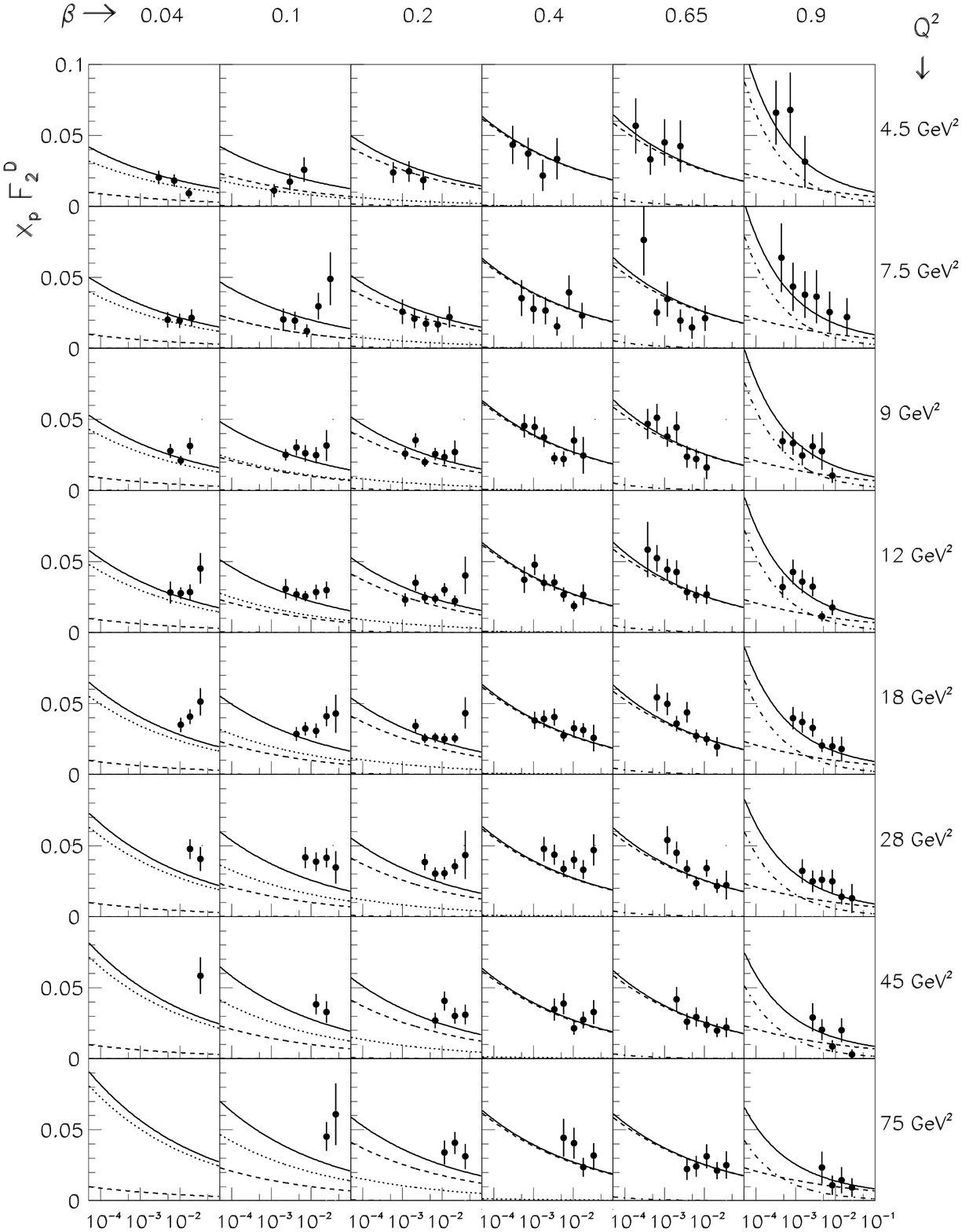,width=10cm}
\caption[]{Fit of two-gluon exchange model to H1 data,
corresponding to softer initial gluons and similar to the ZEUS
fit shown in Fig.~2~\cite{BEKW}.}
\label{fig:quatre}
\end{figure}

We think it is too early to conclude that the data require a singular gluon
structure function in the Pomeron. Further iteration is required to understand
better the different experimental event selections and their modelling. Perhaps
studies of hadronic final states will help resolve these issues~?

\section{Hadronic Final States}

One should always remember that the pseudorapidity $\eta$ (related to the
final-state angle $\theta$) is not the same as the rapidity $y$. The original
ZEUS diffractive event selection was based on a pseudorapidity
cut~\cite{ER}, which tends
to impose a lower cutoff on the virtuality of the quark exchanged between the
$\gamma^*$ and the Pomeron. This tends to bias one away from the ``Pomeron
structure function" region, and to select semi-hard Pomeron events. This effect
should perhaps still be considered in relation to the H1 event selection.
Nowadays, ZEUS base their event selection on a subtraction of the large-$M_x$
background. Several models -- short-range order, triple-Regge, parton
showers~\cite{EGK}
--  suggest that this background should fall off exponentially: $\exp (-b
~M^2_x/W^2)$, where $b\sim 1$ to 2. ZEUS fits this exponent to their data, and
then subtracts the background. This procedure may give problems in the
largest-$M_x$ bin, but not for lower $M_x$, and should not depend strongly on
$Q^2$. Ideally, one would tag diffractive events with a forward spectrometer,
but the statistics is limited. Nevertheless, this subsample may be used to help
tune Monte Carlos, and will in the future provide more useful information.

Diffractive event shapes were a hot topic at DIS~98, with the key issue being
whether they are similar to $e^+e^-$ events~\cite{difffin}. Theoretically,
there is no
particular reason to expect this, since the pattern of perturbative gluon
radiation in $\gamma^*P\rightarrow \bar qqg$ may differ from that in
$e^+e^-\bar qqg$, there may also be a contribution with the Pomeron coupling
directly to a gluon, and there could be important non-perturbative
effects. H1 reports that their diffractive sample has significantly
lower thrust $T$ and higher sphericity $S$, whereas the ZEUS longitudinal
spectrometer data do not have significantly lower $T$ and higher $S$. However,
the ZEUS data have $T$ and $S$ values not very differerent from H1, and they
find that this difference is reduced if they imitate the H1 event selection.

In addition to perturbative contributions, $T$ and $S$ could receive
contributions from non-perturbative remnants of the $\gamma^*$ and/or the
Pomeron. Theorists should make mor detailed calculations of these and other
effects on event shapes, and experimentalists should devise more selective
searches for these possible remnants~\cite{RS}.

Another topic of interest here was the forward-jet cross section, proposed as a
probe of BFKL dynamics. 
ZEUS has reported an excess~\cite{ZEUSfjet} compared to DGLAP-based
models, though the
colour-dipole ARIADNE model does describe them.
A na\"\i ve parton-level BFKL
prediction lay far above the data. However, it has been pointed out that
phase-space corrections reduce significantly the na\"\i ve BFKL parton-level
prediction~\cite{Stirling}, as shown in Fig.~5. Parton-hadron corrections
are small in
ARIADNE, but are difficult
to evaluate in the BFKL picture. One of the problems here is to implement
correctly the cancellations between virtual and real corrections, in the
absence of $k_T$ ordering and incorporating correctly the kinematic constraints
on the gluon $k_T$. Perhaps there will be progress on this before DIS~99~?

\begin{figure}
\hglue4.5cm
\epsfig{figure=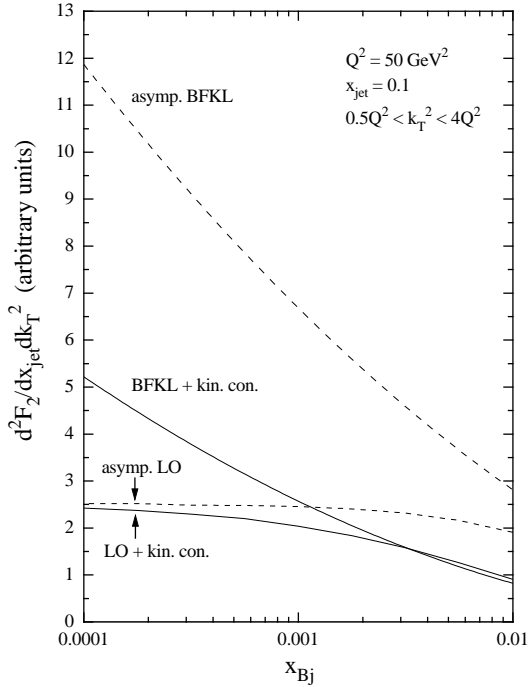,width=7cm}
\caption[]{Calculations of forward jet production at HERA,
showing that an improved BFKL calculation may agree better with the
data~\cite{Stirling}.}
\label{fig:cinq}
\end{figure}

Searches for BFKL effects have also been active in $\bar pp$ collisions at
Fermilab~\cite{FNALBFKL}. Here there is lots of phase space, and an
azimuthal angular
decorrelation between high-$p_T$ jets is a promising signature. However, the
probabilities of rapidity gaps between two high-$p_T$ jets at fixed
$\Delta\eta\geq 1$ has the disappointing feature that
\beq
P^{\rm gap} (630~{\rm GeV}) \sim 3 P^{\rm gap} (1800~{\rm GeV})
\label{nineteen}
\eeq
This may mean that asymptopia has not yet been reached. Fortunately, there is
sufficient kinematic range at Fermilab to extend the study to larger
$\Delta\eta$. If this fails, one can always look forward to the
LHC~\cite{OS}, the
ultimate accelerator for studying diffraction~\cite{FELIX}.

\section{Polarized Structure Functions}

Some impressive new data sets were presented at this meeting. The HERMES
collaboration has presented new data on $g_1$, as well as on hadronic final
states that enable the valence and sea polarized-quark distributions $\Delta
u_V$, $\Delta d_V$ and $\Delta$(sea) to be extracted~\cite{HERMES}. The SMC
has presented their new definitive data on $g^P_1$ and $g^D_1$~\cite{SMC}.
The E155
collaboration~\cite{E155} has presented preliminary data on $g^{P,D}_1$
with very small
statistical errors:
$$
\Delta_{stat} \epsilon t^{0.9}_{0.014} dx~g_1^{P,D}(x) \sim 0.002
$$
Here the challenge will be to control the systematic errors down to a
comparable level. Questions were raised about the nuclear corrections to the
assumed superposition of $D$ and $^4He$ components in the
$^6Li$ target, but in my view the QCD evolution and the small $x$ extrapolation
will be larger issues for attempts to extract $\Gamma_1^{p,D} \equiv
\int^1_0 
dx~g_1^{p,D}(x)$ from these data~\footnote{There have also been
interesting, more precise data on $g_2$~\cite{E155}.}.

The new SMC numbers for these quantities are~\cite{SMC}:
\beq
\Gamma_1^p (10 ~{\rm GeV}^2) = 0.120 \pm 0.005 \pm 0.006 \pm 0.014
\label{twenty}
\eeq
\beq
\Gamma_1^D (10 ~{\rm GeV}^2) = 0.019 \pm 0.006 \pm 0.003 \pm 0.013
\label{twentyone}
\eeq
where the theoretical errors associated with QCD evolution and extrapolation
are listed last. The central value (\ref{twenty}) has been decreased somewhat
by a recalibration of the muon-beam polarization. The CERN data are quite
compatible with the Bjorken sum rule~\cite{SMC}:
\beq
\int^1_0 dx~\left[ g^p_1(x,Q^2) - g^n_1 (x,Q^2)\right]_{Q^2=5{\rm GeV}^2} =
0.173^{+0.024}_{-0.012}
\label{twentytwo}
\eeq
However, the SMC data alone indicate strong discrepancies with the na\"\i ve
singlet sum rules: $\Gamma^p_1$ is off by 3.1$\sigma$, $\Gamma^D_1$ by
3.5$\sigma$. The new SMC results may be combined with previous
data to extract~\cite{EK} a new world average value of the singlet matrix
element:
\beq
\Delta\Sigma = 0.27 \pm 0.05
\label{twentythree}
\eeq
Note that the consistency between the different data is much
improved~\cite{EK} by
including the higher-order perturbative corrections, as seen in Fig.~6.

\begin{figure}
\hglue2.5cm
\epsfig{figure=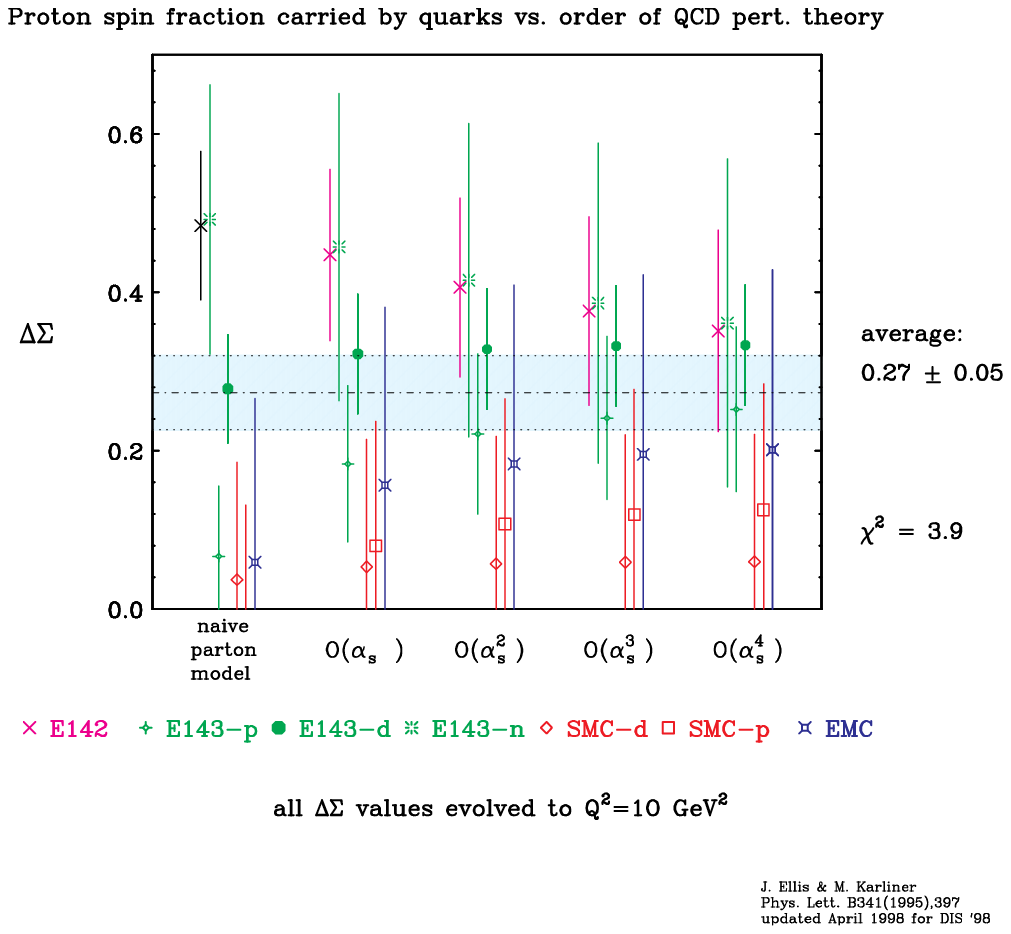,width=12cm}
\caption[]{Values of the total quark contribution to the
proton spin extracted from various sets of data on polarized
lepton-nucleon scattering, showing the improved consistency as
higher-order perturbative QCD corrections are included~\cite{EK}.}
\label{fig:six}
\end{figure}

The SMC has presented here NLO perturbative 
QCD fits to their new data and
those previously available~\cite{NLO}. Independent moment-space and
$(x,Q^2)$-grid
fitting programs give very similar results. They have compared two fashionable
renormalization schemes: the conventional $\overline{\rm MS}$ scheme and the
so-called Adler-Bardeen (AB) scheme in which a gluonic correction is subtracted:
\beq
\widetilde{\Delta q}\vert_{AB} = \Delta q - {\alpha_s\over 2\pi}~\Delta G
\label{twentyfour}
\eeq
These schemes should yield similar $g^{p,D}_1(x,Q^2)$ and $\Delta G(x,Q^2)$,
though they will of course yield different $\Delta q(x,Q^2)$.
The fit values of $\Delta G$ are 
\beq
\overline{\rm MS} : \Delta G = 0.25^{+0.29}_{-0.22} \pm ?
\label{twentyfive}
\eeq
\beq
{\rm AB} : \Delta G = 0.99^{+1.17 + 0.42 + 1.43}_{-0.31 - 0.22 - 0.45}
\label{twentysix}
\eeq
Note that the systematic errors have not yet been evaluated in the
$\overline{\rm MS}$ scheme. There are indications that $\Delta G$ may be
positive, though the two determinations (\ref{twentyfive}),(\ref{twentysix})
are compatible with zero at the 1(2)-$\sigma$ level. Moreover, it is important
to note that, although the central values in 
 (\ref{twentyfive}), (\ref{twentysix}) look rather different, they are
compatible with each other at the 1-$\sigma$ level. Encouragingly, the two fits
give rather similar values of the singlet axial-current matrix element:
\beq
\overline{\rm MS} : a_0 = 0.19 \pm 0.05 \pm 0.04
\label{twentyseven}
\eeq
\beq
{\rm AB} : a_0 = 0.23 \pm 0.07 \pm 0.19
\label{twentyeight}
\eeq
Each of the values (\ref{twentyseven}), (\ref{twentyeight}) is also quite
compatible with the estimate (\ref{twentythree}) based on the moments
$\Gamma_1^{p,n}$. Therefore the proton spin puzzle is still with us.

However, we still do not know whether (in the AB scheme) it is due to $\Delta s
< 0$ and $\Delta G \sim 0$, or to $\Delta s \sim 0, \Delta G > 0$. The need to
measure $\Delta G$ is as great as ever, and even higher on the experimental
agenda. Some insight into this may come from including E155 data in a global
fit, but there will still be concerns about the systematic errors in combining
data from different experiments. It might be possible to obtain some
information from $Q^2$ variations within the E155 data set. Their 10.5$^o$
spectrometer data have yet to be analyzed, though they have less statistics
than the lower-angle, lower-$Q^2$ data. The first direct information on $\Delta
G$ should come from COMPASS~\cite{COMPASS}, measuring the $\bar cc$ (and
possibly charged
hadron-pair) production asymmetry in polarized $\mu N$ scattering:
\beq
\delta (\Delta G/G) \sim 0.1~{\rm to}~ 0.05
\label{twentynine}
\eeq
starting in the year 2000, and from the polarized RHIC
option~\cite{PRHIC},
measuring $\gamma +$ jet and dijet production asymmetries:
\beq
\delta (\Delta G/G) \sim 0.1~{\rm to}~ 0.03
\label{thirty}
\eeq
at $x = 0.02$ to 0.3, also in the same year. The interesting E156 proposal to
measure $\bar cc$ production in polarized eN scattering at SLAC has not been
approved, to my personal regret. This would be a fitting continuation of the
outstanding SLAC polarization programme, and I would like to be convinced if
there are urgent particle-physics arguments why it should not be accepted. 

The
ultimate weapon for determining the polarized gluon distribution would be
the
measurement of dijets, $Q^2$ evolution, etc., using polarized $e$ and $p$ beams
at HERA~\cite{PHERA}. This would be unequalled for measurements at low $x$
and high $Q^2$,
for evaluating sum rules, etc., but may have to wait until beyond 2005.

\section{Large $x$ and $Q^2$}

We have seen at this meeting that the conventional perturbative QCD DGLAP 
evolution describes deep-inelastic data perfectly (almost) everywhere. A global
fit~\cite{MRST} yields
\beq
\alpha_s (M_Z) = 0.1175 \pm 0.003 \, \pm \, ?
\label{thirtyone}
\eeq where the question mark reflects my perplexity, in view  of the fact that
different data sets favour different central values of $\alpha_s$: e.g.,
relatively small for BCDMS and relatively large for SLAC and CCFR. Moreover,
$\gamma/Z^0$ interference is seen clearly at HERA~\footnote{How long will it be
before H1 and ZEUS offer us interesting determinations of $\sin^2\theta_W$?}.
These successes give us confidence that the error of extrapolation to large $x$
and $Q^2$ at HERA is $\leq$ 7 \%.

Is there an experimental excess~\cite{largeQ2} at large $Q^2$?  The
answer is yes, everywhere, both in
neutral currents and in charged currents, both in ZEUS and H1 data, as seen in
the Table.

\begin{center}
\begin{tabular}{|l|l|l|}\hline
&&\\
HI & $Q^2 >$ 15000 GeV$^2$:\phantom{xx} 22 vs 15$\pm$2 & 
 $Q^2 >$ 7500 GeV$^2$: ~~41 vs 28$\pm$8 \\
& M = 200$\pm$12.5 GeV: ~8 vs ~3$\pm$${1\over 2}$ &
 $Q^2 >$ 15000 GeV$^2$: ~9 vs ~~5$\pm$3 \\
&&   \\ \hline
&&\\
ZEUS & $Q^2 >$ 15000 GeV$^2$: 20 vs 17$\pm$2 & 
 $Q^2 >$ 15000 GeV$^2$: 8 vs 4 \\
&$Q^2 >$ 35000 GeV$^2$: ~2 vs 0.3$\pm$0.02 &
 $Q^2 >$ 20000 GeV$^2$: 0.3 vs 1  \\
&& $Q^2 >$ 30000 GeV$^2$: 1 vs 0.06  \\
&&   \\ \hline
\end{tabular}
\end{center}

\vspace*{0.5cm}\noindent 
None of these excesses is significant by itself, and they are probably not
significant when all taken together. Nevertheless, these numbers are
intriguing. Please do not listen too closely to those who tell you the
previous
excesses have now gone away. There are still excesses, and even the much-maligned
1997 data do not exhibit any deficits.

We have all had a lot of fun with the large $Q^2$ and $x$ data during the past
year~\cite{fun}, which also taught the community a great deal. We learnt
how to compile
constraints from HERA and other experiments on leptoquarks and on squarks with
R-violating interactions. Many new analyses at
LEP and
Fermilab~\cite{FNAL} were stimulated: the latter, in particular, make life
almost impossible
for leptoquarks coupling to the first or second generations. On the other hand,
there is still scope for a third-generation leptoquark, or for an R-violating
squark (which may in general have competing R-conserving decays into jets and
missing energy). We have also learnt to compile constraints on contact
interactions, ranging from atomic-physics parity violation upwards in energy.
New analyses of Drell-Yan lepton-pair production at Fermilab~\cite{FNAL} 
and of
$e^+e^-~\rightarrow~\bar ff$ at LEP have also been stimulated. 

Very possibly, the large $x$ and $Q^2$ events are not harbingers of new
physics
beyond the Standard Model. However, there is still plenty of phase space to be
explored in future runs, starting with the expected forthcoming increase in
$e^-p$ luminosity, and continuing with the luminosity upgrade from 2000
onwards.

There was discussion here of HERA events with isolated leptons and missing
$p_T$~\cite{isolated}. H1 reports five $\mu^\pm$ and one $e^-$ event,
as seen in Fig.~7. Four of the $\mu^\pm$ events and the
$e^-$ event resemble kinematically $W^\pm$ production, whereas the remaining
$\mu^\pm$ event looks more like heavy-flavour production: however, H1 expects
the rate for this background to be very small. ZEUS reports $4 e^\pm$ events
that look (more or less) like $W^\pm$ production, compared with expectations of
3.5 $\pm$ 0.4 $e^\pm$ and 1.3 $\pm$ 0.2 $\mu^\pm$ events from $W^\pm$
production and  backgrounds. Twenty years after high-energy $ep$ colliders were
proposed, the HERA experiments seem finally to be detecting the $W^\pm$.

\begin{figure}
\hglue3cm
\epsfig{figure=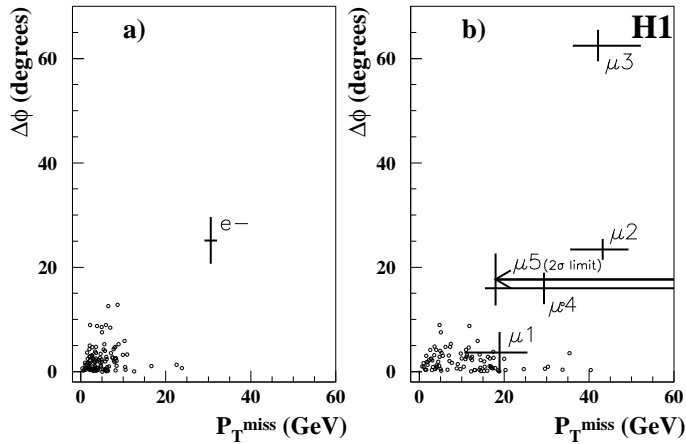,width=10cm}
\caption[]{Scatter plot of isolated lepton + missing transverse
energy events reported by H1, compared with calculations of Standard
Model sources~\cite{isolated}.} 
\label{fig:sept}
\end{figure}

\section{Future Prospects}

There is an active programme of experiments in the years ahead. HERA will
operate with an $e^-$ beam and hopefully at a larger luminosity in 1998
and 1999.
There is a major luminosity upgrade planned to start in 2000. Run~II of the
Fermilab Tevatron collider will start in 2000, and a subsequent Run~III at
higher luminosity is planned to last until the LHC comes on the horizon.
COMPASS~\cite{COMPASS} and polarized RHIC~\cite{PRHIC} will start taking
data in 2000. 

Then the LHC is
scheduled to start taking data in 2005, with the capability to make $pp$, $pA$
and $AA$ collisions, where the nuclei $A$ range from Calcium to Lead.
The question is whether its unique kinematic coverage down to low $x$ and/or up
to large $Q^2$ will be fully exploited. The presently-approved
detectors, ALICE, ATLAS, CMS and LHC-B, are not optimized for QCD studies such as
diffraction. A letter of intent for an additional detector aimed at these
topics has been proposed~\cite{FELIX}, but has not been accepted. This was
not a judgement
on the physics $per~se$, but rather on the small size of the experimental
community expressing initial support for FELIX, and concerns about the
availability of the resources needed for the ambigious FELIX proposal. In view
of the close physics overlap between many of the subjects studied at HERA and
debated hotly at this conference, many of you in the HERA community may wish to
consider some LHC initiative along these lines.

HERA itself has some intriguing future options beyond the luminosity upgrade
and 2005. One of these is to accelerate nuclear beams. This would address many
of the interesting issues in low-$x$ and high parton-density physics, and its
complementarity with the possible LHC programme should be considered carefully.
There are also strong physics motivations for polarizing the HERA proton beam
on a similar time scale, and enthusiasts are developing this physics
case~\cite{PHERA}. The main
hurdle to be overcome in this case is the complication and expense of
accelerating a polarized proton beam: here the operation of polarized RHIC will
provide valuable insight.

In the longer run, there are several interesting options for next-generation DIS
machines~\cite{futureDIS}.
These include LEP $\otimes$ LHC at 67 GeV $\otimes$ 7 TeV, TESLA $\otimes$ HERA
at some 500 GeV $\otimes$ 1 TeV, a muon collider with the Tevatron at some 200
GeV $\otimes$ 1 TeV, and even a proposal to use LEP-like components for
collisions with a booster for the VLHC at 80 GeV $\otimes$ 3 TeV. These
projects would all offer lepton-proton collisions at around 1 TeV in the centre
of mass, carrying significantly further the physics of HERA.

Some of these projects are closer to realization than others. CERN has promised
to take no decisions that could preclude re-installing LEP components in the
LHC tunnel and realizing $ep$ collisions. DESY has chosen the axis of the TESLA
linear collider so that it is tangential to the HERA ring. However, it is not
planned to include $ep$ collisions in the initial TESLA project proposal. A
workshop is currently being organized by ECFA and DESY to explore and develop
the physics case for TESLA. The time is approaching when the physics case for
the $ep$ option should also be developed.

In my view, in order to convince the rest of the physics community and our
political masters to support the major investment required for such a
next-generation collider, you will need more strong arguments than the Pomeron,
low-$x$ physics and diffraction, as was the case for the successful HERA
project proposal. If you want one of these long-term projects to be approved,
or if you favour one of the medium-term post-2005 options for HERA and/or the
LHC, now is the time to build the physics case. We have all enjoyed an
interesting few days here, filled with new experimental measurements and
theoretical results. The next steps are to harness  our
enthusiasm and communicate it to a wider audience.

{\bf Acknowledgements}: It is a pleasure to thank my collaborators Jochen
Bartels, Klaus Geiger, Marek Karliner, Henryk Kowalski, Graham Ross, Jenny
Williams and Mark W\"usthoff for sharing their insights with me: I also
thank Jenny for her help in preparing this talk and Mark for his comments
on the write-up. It is also a pleasure to thank the organizers for their
encouragement and assistance, as well as for their efficient organization
of such a fascinating meeting.

\end{document}